\documentclass{llncs}

\usepackage[english]{babel}
\usepackage{combelow}
\usepackage{xcolor}
\usepackage{eurosym}
\usepackage[hyphens]{url}
\usepackage{graphicx}
\begin{document}

\title{Technological Platform for the Prevention and Management of Healthcare Associated Infections and Outbreaks}
\titlerunning{Technological Platform for HAI Prevention and Management}

\author{Maria Iuliana Bocicor\inst{1} \and Maria Dasc\u{a}lu\inst{2} \and Agnieszka Gaczowska\inst{3} \and Sorin Hostiuc\inst{4} \and Alin Moldoveanu\inst{2} \and Antonio Molina\inst{5} \and  Arthur-Jozsef Molnar\inst{1} \and Ionu\cb{t} Negoi\inst{4} \and Vlad Racovi\cb{t}\u{a}\inst{1}}

\institute{SC Info World SRL, Bucharest, Romania
\and NZOZ Eskulap, Skierniewice, Poland
\and Carol Davila University of Medicine and Pharmacy, Bucharest
\and Polytechnic University of Bucharest, Romania
\and Innovatec Sensing\&Communication, Alcoi, Spain
\email{\{iuliana.bocicor,arthur.molnar,vlad.racovita\}@infoworld.ro, maria.dascalu@upb.ro, \{agaczkowska,soraer,negoiionut\}@gmail.com, alin.moldoveanu@cs.pub.ro, amolina@innovatecsc.com}
}

\maketitle

\begin{abstract}
Hospital acquired infections are infections that occur in patients during hospitalization, which were not present at the time of admission. They are among the most common adverse events in healthcare around the world, leading to increased mortality and morbidity rates, prolonged hospitalization periods and considerable financial burden on both hospitals and patients. Preventive guidelines and regulations have been devised, however compliance to these is frequently poor and there is much room for improvement. This paper presents the prototype of an extensible, configurable cyber-physical system, developed under European Union funding, that will assist in the prevention of hospital infections and outbreaks. Integrating a wireless sensor network for the surveillance of clinical processes with configurable monitoring software built around a workflow engine as key component, our solution detects deviations from established hygiene practices and provides real-time information and alerts whenever an infection risk is discovered. The platform is described from both hardware and software perspective, with emphasis on the wireless network's elements as well as the most important software components. Furthermore, two clinical workflows of different complexity, which are included in the system prototype are detailed. The finalized system is expected to facilitate the creation and automated monitoring of clinical workflows that are associated with over 90\% of hospital infections.
\end{abstract}



\section*{Introduction}
Hospital acquired infections (HAI), also known as healthcare associated infections or nosocomial infections are among the most common adverse events in healthcare around the world, affecting between 8\% and 12\% of patients admitted to hospitals in the European Union \cite{eu_memo_08,collins08}, and 5\% to 10\% of hospitalized patients in the United States every year \cite{cdc}. Hospital infections refer to those infections that occur in patients during hospitalization, which were not present or incubating at the time of admission, including \emph{''infections acquired in the hospital but appearing after discharge, and also occupational infections among staff of the facility''} \cite{WHO02}.

HAI are prevalent across the globe, regardless of geographical, social or economic factors \cite{tikhomirov87,coello93,WHO10,ecdc15,WHO11}. In addition to their most critical consequences, which are an increased rate of mortality and morbidity (an annual number of 37 000 deaths in Europe, 99 000 attributable deaths in the USA \cite{WHO11}, 8 000 in Canada \cite{canada14} and between 7\% and 46\% attributed mortality rates in Southeast Asia \cite{ling15}), HAI also translate in prolonged hospitalization periods (5 to 29.5 days worldwide \cite{WHO11_1}) and further administered treatments. This leads to a considerable financial burden, estimated between \$28 and 45 billion in the United States \cite{stone09}, \$129 million in extra costs incurred in Canada \cite{canada13} \cite{canada14}, and \euro 7 billion in Europe, in 2008 \cite{easac09}. Furthermore, the problem of HAI is closely connected with antimicrobial resistance, another fundamental difficulty for modern healthcare and a significant threat to public health. Many bacterial species, such as Pseudomonas, Acinetobacter, Staphylococcus aureus or Clostridium difficile have developed to be resistant to a wide range of antibiotics. In many cases, they also represent major causes of HAI \cite{mehrad15}.

Having been recognized as a significant issue for 30 years now \cite{tikhomirov87}, surveillance has been in place and measures and precautions have been taken \cite{cdc2016,WHO15} especially during the past decade, in order to reduce nosocomial infection rates. When specific steps are taken by medical personnel, infection rates have been shown to decrease by more than 70\% \cite{cdc2016}. Preventive efforts have proven to be successful, as a report from the Agency of Healthcare Research and Quality \cite{ahrq13} estimates a 17\% decline in the rate of HAI from 2010 to 2013 in the United States. However, the issue is still a delicate one and there is much room for improvement.

This paper presents an extensible, configurable cyber-physical system that will assist in the prevention of HAI and outbreaks \cite{enase17}. Integrating a wireless sensor network (WSN) for the surveillance of clinical workflows with configurable monitoring software, our solution detects deviations from established hygiene practices and provides real-time information and alerts, whenever nonconformity is discovered. The hardware network of wireless sensors that can be deployed within variable-sized clinical locations collect real-time information from the clinical setting, such as substance or material availability (soap, antimicrobial agents, sterile gloves) and environmental conditions affecting the spread of pathogens (oxygen levels, airflow, temperature), thereby providing a complete image of the hospital environment in real time. Monitoring of complex processes such as management of indwelling urinary catheters, postoperative care, intubation or endoscopy is possible by describing them using software workflows that are interpreted and executed by a workflow engine. When the sequence of transitions inferred by the system from sensor data presents deviations from the expected flow, the system alerts responsible personnel. The system will also provide advanced analytics, which are extremely important for pinpointing difficult infection sources that elude existing workflows and for the identification of existing activities targeted towards outbreak prevention and control.

The system is researched and built within a European Union-funded project and its current stage of development represents a proof of concept, including multifunctional smart sensors for monitoring the use of soap, antimicrobial gel and water sink together with two clinical workflows that describe the required hygiene procedures in the case of the general practitioner's office and for minor surgeries that are performed within the clinic where the system will be first deployed. This paper details the smart devices employed, the hardware-software integration as well as the software components that ensure the cyber-physical system achieves its objective of lowering the number and severity of hospital infections.

\section{State of the Art}

In recent years several concrete measures have been taken to reduce the risk of infections in hospitals. As such, guidelines and rules of prevention have been devised for healthcare personnel and patients' safety \cite{mehta14}. As a response of the necessities within the hospital environment, various software or hybrid hardware and software systems have been developed to ensure strict compliance with and enforcement of these instructions.

\subsection{Monitoring Hand Hygiene}
Proper hand hygiene is considered the single most valuable tool in preventing the spread of healthcare-associated infections \cite{mehta14}. Despite existing information, hand hygiene measures are still not widely adopted, being applied in only 40\% of cases when it is required \cite{shea08}. Thus, the underlying idea used by quite a large number of systems is continuous monitoring of healthcare workers' hand hygiene and real-time alert generation in case of non-compliance with established guidelines.

Numerous existing Information and Communication Technology (ICT) solutions use this idea of continuous surveillance and immediate notification in case of hygiene rule violation. IntelligentM \cite{intelligentM} and Hyginex \cite{hyginex} are two systems that combine sensors attached to soap, disinfectant dispensers or faucets with Radio Frequency Identification (RFID) enabled bracelets designed to be worn by medical personnel. Whenever hygiene events are omitted, the systems detect the violation and the bracelet alerts the clinician using vibrations or luminous signals. Biovigil Technology \cite{biovigil} and MedSense \cite{medsense} are similar systems which use wearable badges attached to healthcare workers' uniforms. While Biovigil uses chemical sensors placed at hospital ward entrances to detect whether hand hygiene is undertaken, MedSense employs beacons placed above patient beds or in other points of care, with the aim to establish a wireless patient zone. This allows the badge to identify hand hygiene opportunities by detecting when the badge enters or exits a patient zone. SwipeSense \cite{swipe_sense} is composed of a series of small, recyclable alcohol-based gel dispensers, which can be worn by medical personnel together with proximity sensors mounted on hospital walls and a monitoring web platform. Due to this design, the system allows clinicians to perform hand hygiene without interrupting their activities to walk to a sink or a disinfectant dispenser \cite{simonette13}. UltraClenz's Patient Safeguard System \cite{ultraclenz} is somewhat similar to MedSense, in that it employs a patient based approach, as opposed to the room based approach of previously presented technologies. The system prompts workers to sanitize before and after every patient contact. Unlike other systems presented so far, DebMed \cite{debmed} does not use RFID technology, nor any devices for  medical personnel. It integrates a wireless network of dispensers that send data to a server via hubs and modems installed on each floor. The server application uses a customizable algorithm \cite{diller14} to estimate the number of hand hygiene opportunities per patient-day and compares this number with the actual hand hygiene events that were performed.

\subsection{Disinfection Robots}
The systems mentioned in the previous section prevent the spread of HAI by ensuring that transmission is not induced by medical personnel via contaminated hands. But transmission can also occur via air or contaminated surfaces. Preventing this using chemicals or ultraviolet light has proven successful.

Short-wavelength ultraviolet (UV-C) light provides a strong germicidal effect, by inducing cellular damage and cell death in pathogens. Water, air and surfaces can be purified using UV-C. However, humans must avoid direct UV-C irradiation due to its harmful effects. 

Several types of solutions using UV-C or chemical substances have been developed for air and surface disinfection, in the form of disinfection robots. Tru-D Smart UVC \cite{trudi} scans the room to be disinfected and computes the optimal UV-C light dose required for disinfection according to the particularities of the room. These include its size, geometry, surface reflectivity and the amount and location of equipment present. The robot performs disinfection of the entire room, from top to bottom in one cycle and from one location, ensuring that the ultraviolet light reaches even shadowed areas. The Xenex ''Germ-Zapping Robot'' \cite{xenex} called ''Little Moe'' can disinfect a room using pulses of high-intensity, high-energy ultraviolet light. The deactivation of pathogens takes place in less than five minutes and the disinfected room remains at a low microbial load until it is re-contaminated by a person or the ventilation system. The UV-Disinfection Robot developed by Blue Ocean Robotics \cite{blue_ocean} has the same purpose as Little Moe and Tru-D. The robot can drive autonomously when called upon by medical personnel and the approximate time it needs for fully disinfecting a room is between 10 to 15 minutes.

In addition to UV-C light, certain chemical substances or chemical reactions can be used to eliminate harmful bacteria. The Bioquell Q-10 robot \cite{bioquell} emits hydrogen peroxide vapours, which are safe for hospital and equipment surfaces, as well as for other technological machinery and computers, but are deadly to pathogens. This antibacterial bleaching agent is also toxic to humans and therefore another solution must be distributed across the room after disinfection, to make it safe for humans to enter. A different approach is taken by Sterisafe Decontamination technology \cite{sterisafe}: the Sterisafe robot does not employ any chemicals, but disinfects rooms, including corners and shadowed areas, and removes gases and harmful particles using activated oxygen, also known as ozone. Compared to UV-C light robots, Sterisafe's main advantages are that it disinfects hard to reach surfaces (under beds, behind equipment), while as opposed to the Bioquell Q-10 robots, it completely removes the ozone and other by-products at the end of the disinfection cycle, leaving the room safe for people. A comparison study effectuated by Andersen et al. \cite{andersen06} concluded that disinfection with UV-C light is very effective, but it's best used in conjunction with chemical disinfection, to ensure good cleaning of shadowed areas.

\subsection{Infections and Outbreak Management}
The fight against infections is reinforced through other types of systems that were designed for infection management, such as clinical decision support systems or identification of models that shape the spread of disease. Protocol Watch \cite{protocol_watch} is a decision support system used to improve compliance with the ''Surviving Sepsis Campaign'' international guidelines \cite{surviving_sepsis}, simplifying the implementation of sepsis prevention protocols by regularly checking certain medical parameters of patients. Other relevant software systems developed to enhance treatment policy in case of an infection outbreak are RL6:Infection \cite{rl6} and Accreditrack \cite{accreditrack}. Through proactive monitoring and integration of data obtained from several hospital systems, RL6:Infection helps responsible persons to optimize their initiatives and to make the right decisions concerning infection prevention, based on data collected and presented by the system. Accreditrack was designed to ensure compliance with hand hygiene guidelines, to verify nosocomial infection management processes as well as to provide visibility and transparency for these processes.

Another beneficial endeavour for modelling the spread of disease, for identification of control policies and for ensuring the adoption of correct medical practices is building contact networks \cite{curtis13}. A number of studies have modelled interactions in clinical settings using data collected by wireless sensor networks and electronic medical records, and illustrated that contact network knowledge can be used in preventing and handling hospital infections \cite{friggeri11,hornbeck12,mastrandrea15,stehle11,vanhems13,voirin15}.

The systems presented in this section are effective, but they all address the issue of hospital infections from a singular direction. Systems for monitoring hand hygiene specifically target processes involving disinfection of hands, disinfection robots are useful for disinfecting rooms and equipment, while systems for managing outbreaks are particularly targeted towards that definite goal. The problem of hospital infections and outbreaks however, is a complex one, and we believe a more comprehensive approach will provide a better solution. The platform we propose intends to tackle HAI from several directions, using a two-pronged approach. First, a basic line of defence able to monitor clinical workflows most prone to infection transmission in real time will handle both pathogen-agnostic and pathogen-specific scenarios. Second, an advanced line of defence will be created, that will employ risk maps and will build contact networks using the data gathered as part of the basic line of defence. To the best of our knowledge, our proposed system is the first of its kind to combine a sensor network and software in a cyber-physical platform of the intended versatility. The following sections of the paper describe our proposed system in more detail.

\section{HAI-OPS - Platform Overview}
\label{overview}
The platform we propose is developed within the Hospital Acquired Infection and Outbreak Prevention System (HAI-OPS) research project \cite{hai-ops}, which aims to build a pragmatic, automated solution to significantly decrease overall mortality and morbidity associated with HAI by specifically targeting their most common sources and pathways of transmission. To achieve this, the system will leverage advances in computing power and availability of custom-developed, affordable hardware that will be combined with a configurable, workflow-based software system \cite{haiops2016}.

HAI-OPS will address several clinical and maintenance processes and procedures, such as hand hygiene, catheter management, invasive procedures and surgical care. The system will employ an approach that allows the definition and execution of custom Business Process Model and Notation (BPMN) \cite{bpmn} encoded workflows, which model various clinical and maintenance processes. The cyber-physical platform will be configurable so that it covers differences between clinical unit location and layout, differences in types and specifics of undertaken procedures, as well as variation between existing hygiene guidelines. Furthermore, it will offer interoperability with hospital information systems (HIS) and will allow patient and infection data analysis using risk maps and contact networks. As such, the system will also be geared to help hospital epidemiologists in the fight against infection and outbreaks.

The cornerstone of the proposed system is the detection of events happening within the monitored location which are deemed relevant to infection prevention. For the purposes of the pilot deployment, the system will detect the following event types:
\begin{itemize}
    \item \emph{Presence of persons within the clinical unit}. The system will monitor employees working in clinical or auxiliary positions as well as patients. Person monitoring will be achieved using passive RFID tags embedded in badges. Employees will be provided with permanent customized badges linked to their user profile. Patients will receive a temporary tag when checking into the clinic reception, a mandatory step before any appointment. The use of passive technology allows keeping tags light and inexpensive, while powered elements are embedded in the environment. 
    \item \emph{Person enters or leaves a room}. The passive tags issued to personnel as well as patients, as described above, will be detected by active RFID antennas mounted within the door frames of monitored rooms. This will allow identification of the person as well as determine their direction of movement.
    \item \emph{Person undertakes hand hygiene}. According to the regulations within the partner clinic, hand hygiene can be undertaken using either soap and water, or antimicrobial gel. Both are at the disposal of medical personnel and are placed near the sink within all rooms where clinical activities require it. The system will employ sensor-equipped soap and gel dispensers that transmit an event whenever they are used. Furthermore, the system will employ active RFID installed near the sink to also identify the person who is undertaking hand hygiene. As such, hand hygiene is a high-level event that the software identifies based on several low-level events: proximity to sink area and use of soap dispenser or antimicrobial gel.
    \item \emph{Person equips gloves}. Dispensers for single-use gloves are installed throughout the clinic, near the sink in all rooms where their use might be required. Dispensers will be equipped with sensors that emit an event whenever they are used.
    \item \emph{Patient examination start}. For most workflows, this is the first contact between patient and medical personnel. Patient examinations always take place within a designated area of the room, which usually has a patient bed installed. The system will detect the patient examination event using several low-level events, namely practitioner and patient proximity to the consultation bed as well as thermal imaging data based on an array-type sensor.
    \item \emph{Equipment is used}. Certain procedures require the use of single-use or sterilized equipment. Dispensers are monitored for the single use items. Equipment that can be reused must be sterilized in the autoclave, and witness strips must be employed to prove that the sterilization process was complete. In this case, barcode strips will be printed and affixed to sterilized packages, which will be scanned using a wall-mounted reader when first opened. This will ensure the traceability of the sterilization process for equipment.
    \item \emph{Surface disinfection}. This is undertaken using spray-type disinfectant in recipients. Like in the case of other dispensers, sensors will detect and emit an event when used.
\end{itemize}

\section{Modelling Clinical Processes}
The HAI-OPS platform will monitor various clinical processes defined and encoded as workflows. A workflow engine will create and execute instances of workflows and emit real-time alerts when infection risks are detected, based on the expected succession of actions encoded. Any clinical process can be seen as a sequence of events, conditions and activities and thus modelled using BPMN notation. In this section we describe and model two workflows that are proposed for implementation within the clinical partner of the project, the NZOZ Eskulap \cite{nzoz} outpatient clinic in Skierniwice, Poland.

\subsection{General Practitioner}
\begin{figure*}[!t]
	\centering
    \includegraphics[width=1\textwidth]{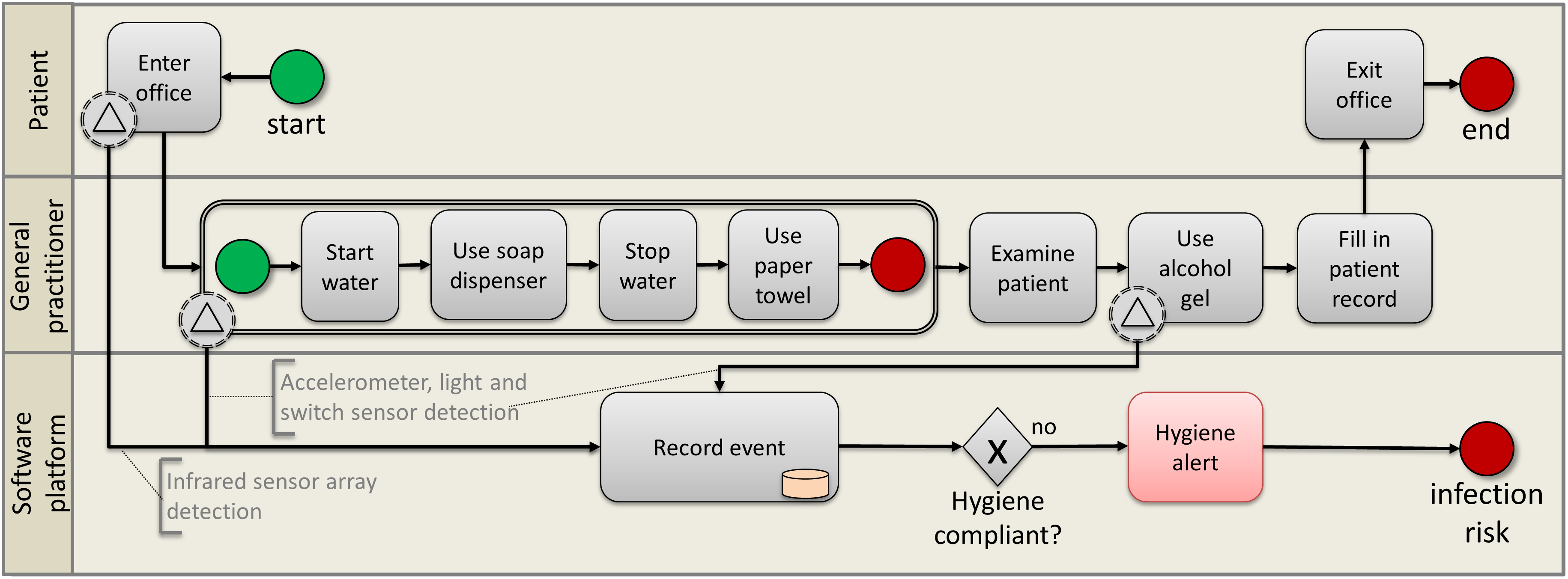}
  	\caption{General Practitioner Workflow (from \cite{enase17})}
  	\label{fig:GPWorkflow}
\end{figure*}

The consultation workflow taking place in the general practitioner (GP) office is the first one considered, given the number of daily consultations and the high degree of event overlap that exists with other workflows taking place within the partner clinic. The simplified BPMN-like model of this workflow, as it takes place within the partner clinic is illustrated in Figure \ref{fig:GPWorkflow}.

An instance of this workflow is created every time a patient enters the GP's office. This is detected using an RFID active system mounted at the door frame, as well as the RFID passive tag the patient receives at the clinic reception when checking in for the appointment. All sensor types employed, along with their positioning inside the GP's office are detailed in Section \ref{wsn}.

After a short conversation, during which the GP might write down various details about the patient's medical issues, patient examination starts. This is detected by a combination of sensors: the RFID tags of the doctor and the patient, a proximity sensor mounted at the side of the bed and additionally an infrared sensor mounted over the bed, used for checking presence. Given the flow modelled in Figure \ref{fig:GPWorkflow}, the system detects that patient examination starts, at which point it checks that hand hygiene was undertaken by the general practitioner. Existing regulations precisely detail how to undertake hand hygiene correctly. Thus, hands must be sanitized according to 10 steps for effective hygiene \cite{who09_2}. 

Assuming that all medical personnel are aware of the detailed actions and correct procedure for proper hand washing, the system will only check that the sink, disinfectants and paper towel dispenser were operated. A collection of sensors mounted on the tap, soap, disinfectant and paper towel dispensers detects the sequence of actions performed. If the sequence is detected in accordance with the provided workflow, the system acknowledges that compliance has been achieved. Otherwise, an alert is  generated, sent to the GP and persisted within the system. In such a situation, if the workflow is violated, its instance is stopped and the hygiene breach is recorded. After receiving notification, the GP will have to perform hand hygiene before examining the patient, or else the system will continue recording and sending alerts. If the workflow continues without interruption, its last step requires the GP to disinfect their hands using antimicrobial gel after the last contact with the patient. This event is again recorded by the system using the same sensors situated in the disinfectant dispenser area. When the examination is finished, the GP updates the patient's record and the patient leaves the office. The workflow is thus completed. All actions detected during  workflow execution are persisted by the system to allow statistics and advanced analyses.

The presented workflow accounts for an ordinary GP consultation. However, some examinations, such as those involving the head, eyes, ears, nose and throat or those concerning patients with skin infections require additional precautions which should be included in the associated workflow. More precisely, new actions must be added, referring to the doctor employing disposable gloves which must be put on before examination and disposed of immediately after the procedure is completed.

\subsection{Minor Surgery}
Minor surgeries represent the most complex clinical workflow currently undertaken within the NZOZ Eskulap clinic \cite{nzoz}. Given the documented risk of infection following surgical procedures, as well as the complexity of invasive procedures that involve both medical personnel and equipment, we believe these workflows are suitable for assessing the impact and performance of the proposed system. Furthermore, ascertaining that our system does not impose additional overhead for existing processes is equally important to consider.

\begin{figure*}[!t]
	\centering
    \includegraphics[width=1\textwidth]{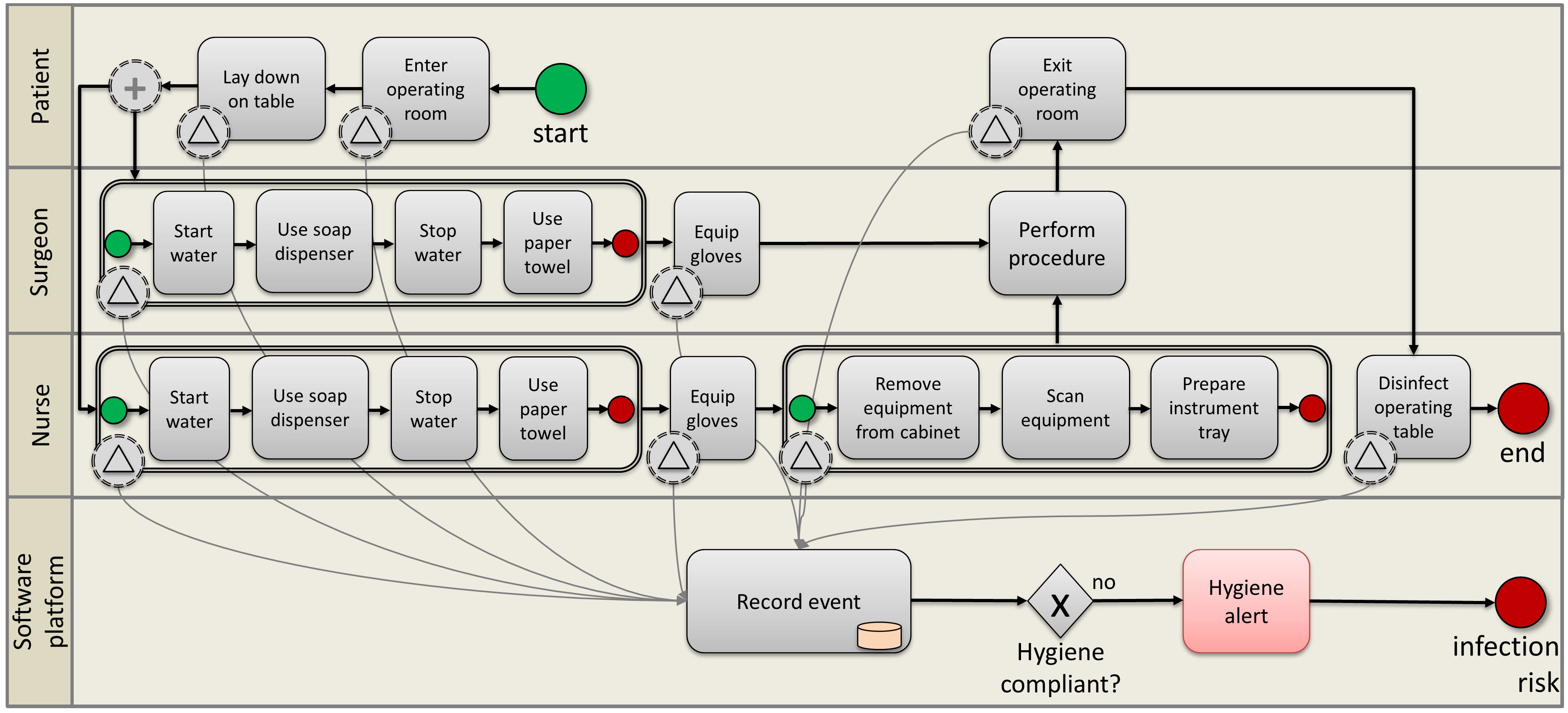}
  	\caption{Minor Surgery Workflow.}
  	\label{fig:MinorSurgeryWorkflow}
\end{figure*}

Minor surgeries are undertaken by a team that consists of a surgeon and a nurse, who must follow a well established hygiene protocol. The equipment used during the procedure is either single use or must be sterilized. A traceability system must be employed to allow proving that equipment underwent proper sterilization. The minor surgery suite consists of a consultation room, which is the first room entered by the patient, and continues with the operating room itself, which can be accessed only from the consultation room. In most cases, minor surgery involves several patient appointments. These include the initial consultation, the intervention itself and one or more follow-up appointments for examination or dressing change. 

Given the purpose of our system, we do not examine the medical procedures that take place in detail, but focus our attention on those steps that experts deem to present high infection risk. In the following we detail the types of surgical appointments that take place at the Eskulap clinic:

\begin{itemize}
    \item \emph{Patient examination}. This is undertaken within the consultation room, in the presence of the surgeon, and with the possible participation of the nurse. Hygiene requirements are similar to those in the general practitioner's office: both the surgeon and nurse must undertake hand hygiene before and after patient examination. 
    \item \emph{Dressing change}. This takes place in the operating room, in the presence of at least one medical practitioner. The hygiene requirements are similar to those during patient examination.
    \item \emph{Minor surgery}. Interventions can take place with or without an immediately preceding patient examination. Surgeries are undertaken only within the operating room, and consist in a series of well defined steps:
        \begin{enumerate}
            \item Appointment starts when the patient enters the operating room 
			\item Patient lays down on the operating table.
			\item The nurse undertakes hand hygiene and equips single use gloves.
			\item The nurse prepares the equipment to be used. Sterilized equipment is taken out of the cabinet, the autoclave sterilization strip is checked and the pouch is opened on the instrument tray.
			\item The surgeon undertakes hand hygiene and equips single use gloves.
			\item After the procedure is complete, the patient leaves the surgery suite.
			\item Nurse disinfects the surgery table.
			\item The appointment is complete.
		\end{enumerate}
\end{itemize}

Given that procedures for patient examination and dressing change are similar to those in the general practitioner's office, Figure \ref{fig:MinorSurgeryWorkflow} illustrates only the expected event flow taking place within the operating room. As shown in the figure, there exists a degree of overlap between the described workflows. This is both expected, as events such as person movement, hand hygiene and equipment use and sterilization are the cornerstones of preventing infection transmission. Furthermore, a high degree of overlap will allow for the reuse of custom developed devices and will lower the cost of deploying and maintaining the system.

\subsection{Known Challenges}
We identified some challenges regarding the interaction of the system with the clinical processes mentioned in the previous subsections, which might emerge particularly due to the various possible constraints. We present these challenges below, including solutions that have already been identified, or potential solutions that are still being investigated within the project.

Firstly, the natural course of clinical processes must not be disrupted by the HAI-OPS platform. The system should not interfere with normal clinical activities, should achieve minimal overhead on clinical processes, as well as high usability and maximal automation so that manual intervention is required only when a risk is detected. Clinical staff must not experience any other modifications from their usual activities; hygiene activities must be performed identically with alerts sent to practitioners through their smartphones.

The placement of sensors within offices, surgery theaters, or generally, in any hospital room, is a challenge within itself. Different types of sensors will be used to detect proximity, position and movement. This requires setting up an appropriate topology for positioning them that will ensure the accurate detection of monitored workflows. Furthermore, the system must take into consideration various restrictions that exist in medical units, such as radio frequency shielding, influence of electromagnetic radiation or Bluetooth interference with medical devices \cite{veterans_affairs,saraf09,wallin03}. To overcome all these, our system will use communication protocols such as healthcare-targeted Bluetooth Low Energy profiles \cite{omre10}, or even wired communication when necessary together with custom designed intelligent devices.

Finally, the most complex challenge regards the possibility of other persons entering the consultation or surgery rooms during examination or minor surgery. If the new person touches one of the practitioners or the patient, they must perform hand disinfection once again. While identification persons can be made as soon as they enter the room, detecting contact between physician, patient and another person is a very difficult task from a technological standpoint. Our project aims to address this challenge by limiting detection to the area of the consultation bed or surgery table using a proximity sensor array placed over the area, as detailed within Section \ref{wsn}. A definitive solution to this challenge is to trigger the execution of a new workflow once an external person enters the examination/surgery room during consultation or minor surgery. We are still investigating possible solutions which involve the clear detection of contact, to avoid having to trigger a new workflow instance in cases when this is not absolutely necessary. 

\section{The Wireless Sensor Network}
\label{wsn}
The hardware platform consists of a Wireless Sensor Network (WSN) addressed to measure and communicate to the software system information about the workflow related actions that are taking place inside the clinical location. All these events are generated by various types of sensors which have different computational, communication and power requirements. For this reason, we designed two different types of hardware nodes: dummy and smart nodes.

Dummy nodes are small, cheap and plain. They are only able to detect simple, low-level actions and send corresponding events to one of the smart nodes. A dummy node usually integrates a proximity sensor and an accelerometer. The proximity sensor features an interrupt function for range detection up to 200mm. The node sends the triggered events to smart nodes using BLE communication, which is further explained in Section \ref{communication}. Both the sensors and the communication module stand for low current consumption, thus it is possible to power supply the device for several years using a small, coin-type battery.   

Dummy nodes are able to detect \textit{what} action is taking place, but the smart nodes are the ones that complete this information by adding \textit{who} is triggering it, \textit{where} and \textit{when} it took place. In the general practitioner's office, the smart node has to find out which person is involved in every action triggered by the dummy nodes (e.g. washing hands, dispensing soap, and so on). To achieve that goal, it communicates with an RFID reader and a thermal array sensor. The location of the smart node inside the GP's office is delimited by the passive infrared array sensor. In this case, it is placed over the consultation bed. This sensor is able to trigger the event when the patient lies down for consultation and when the doctor gets close to them, which is the trigger for starting the patient examination.

The smart node placed within each room will receive RFID information using two antennas. One of them is placed at the door frame and the other at the water sink. Both antennas have different size and gain, which leads to different distance detection ranges. The antenna placed at the door frame aims to detect all badges worn inside the room by patients, practitioners or other identifiable persons. This antenna must be placed opposite to the corridor in order to avoid false positives. On the other hand, the antenna placed near the sink will have a detection range up to 80 centimeters. This is because this antenna must detect only the badge associated to the person using the sink.

\begin{figure*}[!t]
	\centering
    	\includegraphics[width=.75\textwidth]{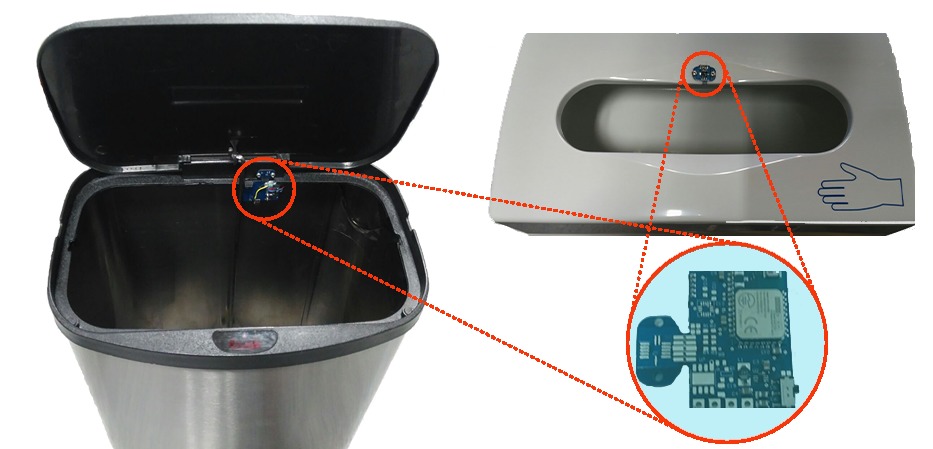}
  	\caption{Dummy nodes located in trash can and gloves dispenser.}
  	\label{fig:sensors}
\end{figure*}

In addition to the smart node, the sensor network in the GP office includes four dummy nodes in charge of triggering low-level events denoting that equipment is being used. These include the soap dispenser, disinfectant gel, disposable gloves and the trash can. All these devices have been modified to electronically detect use and transmit the event to the smart node. Figure \ref{fig:sensors} shows the sensor node with BLE connectivity and its location within the trash can and glove dispenser.

Installation of the sensor network in the minor surgery room is similar to the GP office, except that the minor surgery ward involves two rooms, the consultation room and the operating room. As both include an area for hand hygiene and consultation bed, we find that most devices must be duplicated. RFID readers must be installed at both door rooms and both water sinks. It's also necessary to detect the same triggered events as in the GP office: use of soap dispenser or antimicrobial gel, use of the glove dispenser and trash can. However, there are also some important differences between the surgery room and the GP office. First, there must be a way to ensure that the doctor is going to use disinfected equipment before surgery. After the autoclave sterilization procedure is completed, a barcode sticker is attached to every bag where items are packaged. When the nurse is preparing the equipment, all the items to be used are identified through a barcode reader placed on the wall near the equipment cabinet. Second, surgery lamps prevent thermal array sensors in the ceiling from detecting when the patient is laying down on the operating table. This event is triggered using a combination of infrared sensors and RFID signal quality levels. Finally, the system must detect when the operating table is cleaned after each procedure. For this, a smart holder for the disinfectant spray has been built to detect, using a simple electronic switch, when the spray is used.

\subsection{Communication Infrastructure}
\label{communication}
In June 2010, the Bluetooth Special Interest Group (SIG) published the 4.0 version, which introduced a new specification, known as Bluetooth Low Energy (BLE) \cite{ble}. This allows devices that do not require transmitting large volumes of data to connect to each other and communicate while consuming a minimum amount of energy.

This HAI-OPS platform is developed around this technology. Dummy nodes include a BLE112 Smart Module from Bluegiga \cite{silicon_labs}, which integrates features required for BLE applications including radio, software stack and GATT-based profiles. Moreover, it has flexible hardware interfaces to connect different peripherals and sensors and it is powered directly from a standard 3V coin cell battery. 

The network architecture used in the HAI-OPS platform is a star topology, which in its simplest form consists of a central device that can connect to multiple peripherals. The Generic Attribute Profile (GATT) establishes how to exchange data over BLE. It makes use of the Attribute protocol (ATT) which is used to store services, characteristics and related data in a table. Most of the ATT protocol is pure client-server: the client takes the initiative, while the server responds. But it also has notifications and indication capabilities, in which the server takes the initiative of notifying a client that an attribute value has changed. The lowest concept in GATT communications is the characteristic, which encapsulates single data, such as sensor measurements. 

All data generated by dummy nodes is collected by a smart node and sent to a database server. Smart nodes generate more data than dummy nodes and therefore require a more powerful communication interface. First of all, the smart node must have full time connectivity to the central server and must always be available for receiving events from dummy nodes. Second, data sent by smart nodes has to reach a web server which may be located inside or outside the local area network. In both cases, smart devices have to reach a router or an access point which is most likely located in another room, separated by several walls. Finally, data generated by some sensors like the RFID or the passive-infrared array require a transmission rate of several Mbits/s.
 
The smart node integrates an Ethernet connector which allows it to connect to the local network deployed in the building. In situations when cable installation is not possible in certain locations, the smart device may use its 802.11n wireless module.

Data exchanged between smart nodes and the database server is formatted in JavaScript Object Notation (JSON) format \cite{json}. JSON is text, so it is easy to work with the data as objects, with no complicated paring and translations. Moreover, when storing data, the data has to be a certain format, and text is always one of the legal formats.

\section{Software Components}
The HAI-OPS cyber-physical platform's software architecture employs a client-server model, with a server installed for each clinical unit, to which multiple heterogeneous clients can connect. Considering this aspect, the present section discusses the essential components of the platform's server side, followed by the expected components that make up the client.

Being a cyber-physical platform, the hardware devices required to measure and record activities and their associated software are of major importance. Each such device contains the required networking hardware and software controller that allow it to connect to the HAI-OPS server in order to transmit live data. We refer the reader to Section \ref{wsn} for more details about the hardware side of these devices. The software applications that manage the hardware equipment vary, according to device type: for dummy nodes, we used the ''BLE SW Update Tool'' \cite{silicon_labs}, while smart nodes' software was created using the Python programming language, including support from specialized libraries for operations such as access to system Bluetooth and BLE resources, reading and writing of digital pins, communication with the RFID sensor, Inter-Integrated Circuit communication \cite{i2c} or analog to digital conversion.

\subsection{Server Modules}
Providing real-time alerts in case of detected infection risk is the leading feature of the server software. However, to achieve this, the system first needs to be able to receive data from connected devices and hospital information systems (when available), to process and analyze the data and to monitor the infection risk using workflow technology. In addition, the server-side software will also perform analyses on  collected data, which must thus be persisted. As a consequence of all these, the server main components, which are detailed below, are grouped into four major categories: data acquisition, workflow engine, data store and client facing subsystems.

\subsubsection{Data Acquisition.}
This component's main goal is to send the data recorded by the platform's connected devices to the system's data store persistent repository. A REST architecture \cite{rest} is implemented by the server to receive sensor data and the incoming readings are JSON formatted \cite{json}. All measurements transmitted to the server by the sensors have at least the following essential fields: a Uniform Resource Locator (URL) and a sensor unique identifier for identification of the reading, its source node and the time stamp. In addition to these, each measurement might contain further attributes according to its type: presence sensors might send boolean values, RFID  readers will send the identification tag detected and so on.

As a secondary purpose, the data acquisition component can interact with hospital information systems. Such systems contain a wealth of information that can be used for  prevention of infections and outbreaks, such as patient susceptibility data, arrangement and location of patients and patient beds, information about patients that are immunosuppressed or otherwise at a greater risk for contracting infections. Data received from these external systems is stored in the data repository, from where it is expected to be reused for further analyses.

\subsubsection{Workflow Engine.}
A modelling component allows creating, deleting and updating workflows monitored by the system. These workflows can be executed by any commercial off the shelf workflow engine implementation that understands BPMN notation. The workflow engine interprets events, such as inputs from deployed sensors and acts upon them according to a predefined process, represented by the modelled workflow. The workflow engine component also integrates a generic adapter interface that can be developed to have various implementations, to abstract the particularities of the specific workflow engine employed. As such, HAI-OPS can be used with any major workflow engine implementation, as long as a suitable workflow adapter component is implemented. Monitored workflows can be managed via a user interface, as detailed within section \ref{client} by the system's administrator and they are persisted in the data store.

\subsubsection{Data Store.}
This component is the system’s data repository, being responsible with data persistence for registered users and devices, workflow instances and workflow metadata, as well as raw data recorded from the network of connected devices or any input transmitted by deployed hospital information systems. All stored data will be used for complex analyses part of the system's advanced line of defence geared towards pinpointing elusive reasons of infection and for monitoring outbreaks. The data store is implemented using Couchbase Server \cite{couchbase}.

\subsubsection{Client Facing Subsystems.}
This component includes those subsystems which are connected to client components detailed in section \ref{client}, offering server-side functionality for real-time alerting, data analysis and user and device management.

\textit{Real-Time Alerting.} This is a key component of the system, as it is directly responsible with creating and transmitting the alerts. Whenever an instantiated workflow reaches a point where an infection risk is detected, the workflow engine adapter component will send the required data to this subsystem, which will create an alert and send it to responsible end-users. All alerts contain at least the following information: the workflow, device and person responsible as well as date and time information together with a textual description. The alert data is sent to the user through the alerter client component installed on the users' smartphone. In case of alert, the involved person will have to take corrective measures. All generated alerts are persisted in the data store.

\textit{Data Analysis.} This component provides the advanced data analysis capabilities of the platform. Using data received from the connected devices and existing hospital information systems, its purpose is to aggregate context information and sensor readings and provide information regarding outbreak and infection risk, as well as to facilitate the identification of infection sources and means of transmission in the case of outbreak. It is directly connected to the client epidemiology user interface detailed in section \ref{client}, which allows epidemiologists to visualize analyses results.

\textit{User and Device Management.} All information regarding users and devices is deposited within the data store repository, thus being made available to other system components. Both entity types need to be uniquely identifiable, as workflow execution and alert transmission is tightly linked with involved users and connected devices registered within the system. System users can have one of the administrator, epidemiologist or clinical personnel roles with a role-based access permission system put into place.

\subsection{Client Components}
\label{client}
Two client applications are included in the HAI-OPS platform: the \emph{Alerter Client} mobile application used to transmit alerts to clinical personnel using their smartphones or other smart wearables, and the \emph{Administration Client} web application that  provides the required features to enable management of connected devices, users and workflows.

\subsubsection{Alerter Client.}
Users registered within the system will have the Alerter Client application installed on their smartphone. The application will provide two features: (1) Receive push notification when an alert is generated for the involved person; (2) View history of past alerts for the involved person. The received message will contain detailed information regarding why the alert was generated as well as a meaningful description. As further development, we will also analyze the possibility of using alternate means of notification, such as using short text messages (SMS) that will be received by the registered user’s mobile device, which would allow personnel to receive alerts without installing additional software. While the current platform of choice is the smartphone, the system can work with any programmable device, such as smart wearables or a custom design active badge based on RFID technology.

\subsubsection{Administration Client.}
User, workflow and smart device administration will be achieved using a web application locally installed at each medical unit where the HAI-OPS platform is deployed. The administration client will provide user interfaces for:
\begin{itemize}
    \item Visualizing the complete history and details for the alerts that were generated for a specific user. Users having administrator privileges will be able to view the entire alert history for any registered user.
    \item Managing HAI related data. This interface is targeted towards epidemiologists and will allow them to visualize the results of various statistical analyses applied on collected data (e.g. statistics based on historical alert data, aggregated by user, workflow or by sensor). This information will allow identification of infection and outbreak hotspots, as well as of locations where additional disinfection procedures or staff are needed. This interface will also have access to more complex reports, including risk maps and contact networks which, in conjunction with historical alert data, are important instruments for the identification of transmission pathways.
    \item Administration of connected devices, registered users and monitored workflows. The associated user interface will only be available to users registered as administrators. They are responsible with managing system users and employed smart devices. When the system is deployed in a clinical unit, the administrator will manage the monitored workflows, and will modify them as required, maintaining a continuous communication with the clinical personnel.
\end{itemize}

\section*{Conclusions}
Seven to ten people out of every 100 hospitalized patients worldwide acquire at least one hospital infection \cite{WHO11}. The risk of contacting or transmitting a hospital acquired infection can be greatly reduced, provided that medical units are equipped with efficient tools that ensure compliance to sanitation regulations and that medical personnel pay particular attention to hygiene.

Through our research within the HAI-OPS project \cite{hai-ops} and through the platform under development, we are aiming to bring a contribution towards decreasing infection-related morbidity and mortality, to help prevent outbreaks and also have an indirect positive impact regarding other connected issues, such as the war against antimicrobial resistant pathogens.

The HAI-OPS platform is still under development and the present paper depicts this system in its current stage, emphasizing both hardware and software components. Using a wireless network of smart devices and sensors to monitor clinical processes that might be involved in infection transmission, the system detects potential risks in real-time and immediately alerts involved persons. Two clinical processes of different complexity, which were included in the system's prototype, as they were selected for implementation during the first pilot deployment, are detailed in this paper for exemplification: general practitioner examination and minor surgery. Similarly to all clinical processes that will be included in the system, according to medical activities that take place at various deployment locations, these processes are BPMN encoded and are executed by a workflow engine. In case of a process violation, which is detected via the wireless sensor network, the engine communicates this to the real-time alerting subsystem, which promptly sends notifications to responsible medical personnel. Thus, any clinical process can only be completely and successfully executed when preventive guidelines are followed.

Current achievements constitute the basic line of defense which our system offers for protection against infections. As further development, we intend to add an advanced line of defence, which will bring the platform to its maturity. This will include advanced analysis tools and algorithms to process data collected by sensors during large periods of time. Together with information extracted from hospital systems, risk maps will be constructed to depict the degree of infectious risk at room level. Contact networks will be used to analyze the source and spread of infection. These will be presented in an easy to understand, visual form and will assist epidemiologists in pinpointing elusive reasons of infection, in monitoring outbreaks and, most importantly, in planning infection prevention and control.

\subsubsection*{Acknowledgments.}
This work was supported by a grant of the Romanian National Authority for Scientific Research and Innovation, CCCDI UEFISCDI, project number 47E/2015, \emph{HAI-OPS - Hospital Acquired Infection and Outbreak Prevention System}.

\bibliographystyle{splncs03}
\bibliography{bibliography}

\begin{thebibliography}{10}
\providecommand{\url}[1]{\texttt{#1}}
\providecommand{\urlprefix}{URL }

\bibitem{ahrq13}
{Agency for Healthcare Research and Quality}: {Interim Update on 2013 Annual
  Hospital-Acquired Condition Rate and Estimates of Cost Savings and Deaths
  Averted From 2010 to 2013}.
  \url{https://www.ahrq.gov/sites/default/files/wysiwyg/professionals/quality-patient-safety/pfp/interimhacrate2013.pdf}
  (2013)

\bibitem{andersen06}
Andersen, B., Banrud, H., Boe, E., Bjordal, O., Drangsholt, F.: {Comparison of
  UV C light and chemicals for disinfection of surfaces in hospital isolation
  units}. Infect Control Hosp Epidemiol  27,  729--734 (2006)

\bibitem{bioquell}
{Bioquell}: {Bioquell Q-10}.
  \url{http://www.bioquell.com/en-uk/products/life-science-products/archive-hc-products/bioquell-q10/}
  (2016)

\bibitem{biovigil}
{BIOVIGIL Healthcare Systems, Inc.}: Biovigil and our team.
  \url{http://www.biovigilsystems.com/about/} (2015)

\bibitem{blue_ocean}
{Blue Ocean Robotics}: Uv-disinfection robot.
  \url{https://blue-ocean-robotics.com/uv-disinfection/} (2017)

\bibitem{ble}
{{B}luetooth {SIG}, {I}nc.}: Bluetooth low energy.
  \url{https://www.bluetooth.com/what-is-bluetooth-technology/how-it-works/low-energy}
  (2017)

\bibitem{enase17}
Bocicor, I., Dascalu, M., Gaczowska, A., Hostiuc, S., Moldoveanu, A., Molina,
  A., Molnar, A.J., Negoi, I., Racovita, V.: Wireless sensor network based
  system for the prevention of hospital acquired infections. In: 13th
  International Conference on Evaluation of Novel Approaches to Software
  Engineering (2017)

\bibitem{haiops2016}
Bocicor, M.I., Molnar, A.J., Taslitchi, C.: Preventing hospital acquired
  infections through a workflow-based cyber-physical system. In: Proceedings of
  the 11th International Conference on Evaluation of Novel Software Approaches
  to Software Engineering. pp. 63--68 (2016)

\bibitem{canada14}
{Canadian Union of Public Employees}: Health care associated infections:
  backgrounder and fact sheet.
  \url{http://cupe.ca/health-care-associated-infections-backgrounder-and-fact-sheet}
  (2014)

\bibitem{cdc}
{Centers for Disease Control and Prevention}: {Preventing Healthcare-Associated
  Infections}.
  \url{https://www.cdc.gov/washington/~cdcatWork/pdf/infections.pdf}

\bibitem{cdc2016}
{Centers for Disease Control and Prevention}: {HAI Data and Statistics}.
  \url{https://www.cdc.gov/hai/surveillance/} (2016)

\bibitem{coello93}
Coello, R., Glenister, H., Fereres, J., Bartlett, C., Leigh, D., Sedgwick, J.,
  Cooke, E.: The cost of infection in surgical patients: a case-control study.
  Journal of Hospital Infections  25,  239--250 (1993)

\bibitem{collins08}
Collins, A.: Preventing health care–associated infections. In: Fagerberg, J.,
  Mowery, D.C., Nelson, R.R. (eds.) Patient Safety and Quality: An
  Evidence-Based Handbook for Nurses, chap.~41, pp. 547--570. Agency for
  Healthcare Research and Quality (US) (2008)

\bibitem{couchbase}
{Couchbase}: {Couchbase Server}. \url{https://www.couchbase.com/} (2017)

\bibitem{curtis13}
Curtis, D., Hlady, C., Kanade, G., Pemmaraju, S., Polgreen, P., Segre, A.:
  Healthcare worker contact networks and the prevention of hospital-acquired
  infections. Plos One  (2013), {DOI: 10.1371/journal.pone.0079906}

\bibitem{debmed}
{DebMed - The Hand Hygiene Compliance and Skin Care Experts}: A different
  approach to hand hygiene compliance.
  \url{http://debmed.com/products/electronic-hand-hygiene-compliance-monitoring/a-different-approach/}
  (2016)

\bibitem{veterans_affairs}
{Department of Veterans Affairs}: {MRI Design Guide}.
  \url{https://www.wbdg.org/ccb/VA/VADEGUID/mri.pdf} (2008)

\bibitem{diller14}
Diller, T., Kelly, J., Blackhurst, D., Steed, C., Boeker, S., McElveen, D.:
  {Estimation of hand hygiene opportunities on an adult medical ward using
  24-hour camera surveillance: Validation of the HOW2 Benchmark Study}.
  American Journal of Infection Control  42,  602--607 (2014)

\bibitem{json}
{Ecma International}: {The JSON Data Interchange Format}.
  \url{http://www.ecma-international.org/publications/files/ECMA-ST/ECMA-404.pdf}
  (2013)

\bibitem{easac09}
{European Academies Science Advisory Council}: {Healthcare-associated
  infections: the view from EASAC}.
  \url{http://www.easac.eu/fileadmin/PDF_s/reports_statements/Healthcare-associated.pdf}
  (2013), the Royal Society, London

\bibitem{ecdc15}
{European Centre for Disease Prevention and Control}: Annual epidemiological
  report. antimicrobial resistance and healthcare-associated infections. 2014.
  \url{http://ecdc.europa.eu/en/publications/Publications/antimicrobial-resistance-annual-epidemiological-report.pdf}
  (2015)

\bibitem{eu_memo_08}
{European Commission}: {Questions and Answers on patient safety, including the
  prevention and control of healthcare associated infections}.
  \url{http://europa.eu/rapid/press-release_MEMO-08-788_en.htm} (2008)

\bibitem{accreditrack}
{Excelion Technology Inc.}: Accreditrack.
  \url{http://www.exceliontech.com/accreditrack.html} (2013)

\bibitem{rest}
Fielding, R.T.: Architectural styles and the design of network-based software
  architectures (2000), doctoral dissertation, University of California

\bibitem{friggeri11}
Friggeri, A., Chelius, G., Fleury, E., Fraboulet, A., Mentre, A., Lucet, J.C.:
  Reconstructing social interactions using an unreliable wireless sensor
  network. Computer Communications  34,  609--618 (2011)

\bibitem{medsense}
{General Sensing}: Medsense clear. hand hygiene compliance monitoring.
  \url{http://www.generalsensing.com/medsenseclear/} (2014)

\bibitem{canada13}
{Government of Newfoundland and Labrador. Department of Health and Community
  Services}: {HAI Report 2009-2012}.
  \url{http://www.health.gov.nl.ca/health/publichealth/cdc/hai/hai_2012.pdf}
  (2013)

\bibitem{hai-ops}
{HAI-OPS}: home page. \url{http://haiops.eu} (2017)

\bibitem{hornbeck12}
Hornbeck, T., Naylor, D., Segre, A.M., Thomas, G., Herman, T., Polgreen, P.M.:
  Using sensor networks to study the effect of peripatetic healthcare workers
  on the spread of hospital associated infections. J Infect Dis.  206,
  1549--1557 (2012)

\bibitem{hyginex}
{Hyginex}: Introducing hyginex generation 3. \url{http://www.hyginex.com/}
  (2015)

\bibitem{mastrandrea15}
Mastrandrea, R., Soto-Aladro, A., Brouqui, P., Barrat, A.: Enhancing the
  evaluation of pathogen transmission risk in a hospital by merging
  hand-hygiene compliance and contact data: a proof-of-concept study. {BMC
  Research Notes}  8,  426 (2105)

\bibitem{mehrad15}
Mehrad, B., Clark, N., Zhanel, G., Lynch, J.: Antimicrobial resistance in
  hospital-acquired gram-negative bacterial infections. Chest.  147,
  1413--1421 (2015)

\bibitem{mehta14}
Mehta, Y., Gupta, A., Todi, S., Myatra, S., Samaddar, D., Bhattacharya, P.V.P.,
  Ramasubban, S.: Guidelines for prevention of hospital acquired infections.
  Indian Journal of Critical Care Medicine  18,  149--163 (2014)

\bibitem{ling15}
M.L., L., A., A., G., M.: The burden of healthcare-associated infections in
  southeast asia: A systematic literature review and meta-analysis. Clinical
  Infectious Diseases  60,  1690--9 (2015), dOI: 10.1093/cid/civ095

\bibitem{i2c}
{NXP Semiconductors}: I2c-bus specification and user manual.
  \url{http://www.nxp.com/docs/en/user-guide/UM10204.pdf} (2014)

\bibitem{nzoz}
{NZOZ Eskulap}: {NZOZ Eskulap}. \url{www.eskulapskierniewice.pl/} (2016)

\bibitem{bpmn}
{Object Management Group}: Business process model and notation.
  \url{http://www.bpmn.org/} (2015)

\bibitem{omre10}
Omre, A.: {Bluetooth Low Energy: Wireless Connectivity for Medical Monitoring}.
  Journal of Diabetes Science and Technology  4,  457--463 (2010)

\bibitem{protocol_watch}
{Philips}: Protocolwatch - {SSC Sepsis}.
  \url{http://www.healthcare.philips.com/main/products/patient_monitoring/products/protocol_watch/}
  (2015)

\bibitem{rl6}
{RL Solutions}: {The RL6 Suite / Infection Surveillance}.
  \url{http://www.rlsolutions.com/rl-products/infection-surveillance} (2015)

\bibitem{intelligentM}
Ryan, J.: Medtech profiles: Intelligentm - a simple yet powerful app to
  dramatically reduce hospital-acquired infections.
  \url{https://medtechboston.medstro.com/profiles-intelligentm/} (2013)

\bibitem{saraf09}
Saraf, S.: Use of mobile phone in operating room. Journal of Medical Physics
  34,  101--1002 (2009)

\bibitem{shea08}
{SHEA/IDSA Practice Recommendation}: Strategies to prevent surgical site
  infections in acute care. Infection Control and Hospital Epidemiology  29
  (2008)

\bibitem{silicon_labs}
{Silicon Labs}: {Bluegiga Bluetooth Smart Software Stack}.
  \url{https://www.silabs.com/products/development-tools/software/bluegiga-bluetooth-smart-software-stack}
  (2017)

\bibitem{simonette13}
Simonette, M.: Tech solutions to hospital acquired infections.
  \url{http://www.healthbizdecoded.com/2013/06/tech-solutions-to-hospital-acquired-infections/}
  (2013)

\bibitem{stehle11}
Stehle, J., Voirin, N., Barrat, A., Cattuto, C., Colizza, V., Isella, L.,
  Regis, C., Pinton, J.F., Khanafer, N., Van~den Broeck, N., Vanhems, P.:
  {Simulation of an SEIR infectious disease model on the dynamic contact
  network of conference attendees}. {BMC Medicine}  9 (2011)

\bibitem{sterisafe}
{SteriSafe ApS}: Decontamination and disinfection robot for hospitals.
  \url{http://sterisafe.eu/about-sterisafe/} (2017)

\bibitem{stone09}
Stone, P.: Economic burden of healthcare-associated infections: an american
  perspective. Expert Rev Pharmacoecon Outcomes Res.  9,  417--422 (2009)

\bibitem{surviving_sepsis}
{Surviving Sepsis Campaign}: {International Guidelines for Management of Severe
  Sepsis and Septic Shock: 2012}.
  \url{http://www.sccm.org/Documents/SSC-Guidelines.pdf} (2012)

\bibitem{swipe_sense}
{Swipe Sense}: Hand hygiene. redefined. \url{https://www.swipesense.com/}
  (2015)

\bibitem{tikhomirov87}
Tikhomirov, E.: {WHO} programme for the control of hospital infections.
  Chemioterapia  6,  148--151 (1987)

\bibitem{trudi}
{Tru-D Smart UVC}: About tru-d. \url{http://tru-d.com/why-uvc-disinfection/}
  (2016)

\bibitem{ultraclenz}
{UltraClenz}: Patient safeguard system.
  \url{http://www.ultraclenz.com/patient-safeguard-system/} (2016)

\bibitem{vanhems13}
Vanhems, P., Barrat, A., Cattuto, C., Pinton, J.F., Khanafer, N., Regis, C.,
  Kim, B., Comte, B., Voirin, N.: Estimating potential infection transmission
  routes in hospital wards using wearable proximity sensors. {PloS One}  8,
  e73970 (2103)

\bibitem{voirin15}
Voirin, N., Payet, C., Barrat, A., Cattuto, C., Khanafer, N., Regis, C., Kim,
  B., Comte, B., Casalegno, J.S., Lina, B., Vanhems, P.: Combining
  high-resolution contact data with virological data to investigate influenza
  transmission in a tertiary care hospital. Infection Control \& Hospital
  Epidemiology  36,  254--260 (2015)

\bibitem{wallin03}
Wallin, M., Wajntraub, S.: Evaluation of bluetooth as a replacement for cables
  in intensive care and surgery. Critical Care and Trauma pp. 763--767 (2003)

\bibitem{WHO02}
{World Health Organization}: Prevention of hospital-acquired infections - a
  practical guide.
  \url{http://www.who.int/csr/resources/publications/whocdscsreph200212.pdf}
  (2002)

\bibitem{who09_2}
{World Health Organization}: How to handwash?
  \url{http://www.who.int/gpsc/5may/How_To_HandWash_Poster.pdf} (2009)

\bibitem{WHO10}
{World Health Organization}: The burden of health care-associated infection
  worldwide. \url{http://www.who.int/gpsc/country_work/summary_20100430_en.pdf}
  (2010)

\bibitem{WHO11}
{World Health Organization}: Health care-associated infections - fact sheet.
  \url{http://www.who.int/gpsc/country_work/gpsc_ccisc_fact_sheet_en.pdf}
  (2011)

\bibitem{WHO11_1}
{World Health Organization}: Report on the burden of endemic health
  care-associated infection worldwide.
  \url{http://apps.who.int/iris/bitstream/10665/80135/1/9789241501507_eng.pdf}
  (2011)

\bibitem{WHO15}
{World Health Organization}: {Clean Care is Safer Care - Five moments for hand
  hygiene}. \url{http://www.who.int/gpsc/tools/Five_moments/en/} (2015)

\bibitem{xenex}
{Xenex}: Xenex germ-zapping robots. \url{http://www.xenex.com/} (2015)

\end{thebibliography}

\end{document}